# Pressure-induced phonon frequency shifts in transition-metal nitrides

Xiao-Jia Chen, Viktor V. Struzhkin, Simon Kung[*], Ho-kwang Mao, and Russell J. Hemley

Geophysical Laboratory, Carnegie Institution of Washington, Washington, DC 20015, USA

Axel Nørlund Christensen

Højkolvej 7, DK-8210 Aarhus V Denmark

We report the first experiments on the high pressure phonon spectra of transition-metal nitrides HfN, ZrN, and NbN, obtained by Raman scattering measurements. Two pronounced bands, which are related to the acoustic part at low frequency around 200 cm$^{-1}$ and the optical part at high frequency around 550 cm$^{-1}$ of the phonon spectrum, respectively, shift to high frequency values with increasing pressure. An analysis of the results allows us to reproduce the experimental pressure dependence of the superconducting transition temperature $T_c$ of ZrN and NbN.

[*] Winston Churchill High School, Potomac, MD 20854, USA

## I. INTRODUCTION

When pressure is applied to a superconducting material the superconducting transition temperature $T_c$ generally changes. In contrast to the decrease of $T_c$ generally observed when a nontransition-metal superconductor is subjected to high pressure, many of the transition metals and their alloys and compounds exhibit an increase in $T_c$.[1] The recent development of new high pressure techniques for electric and magnetic measurements in a diamond-anvil cell (DAC) makes possible the investigation of the pressure dependence of $T_c$ at megabar range.[2,3] A significant increase in $T_c$ with pressure up to 120 GPa has been observed in some transition metals,[4,5] pointing to the possibility of obtaining high $T_c$'s at high pressures in their alloys and compounds. Compared to the elements, transition metal carbide and nitride compounds have relatively high values of $T_c$ at ambient condition, reaching nearly 18 K in $NbC_{0.3}N_{0.7}$.[6] Earlier studies[7-10] have revealed that application of pressure increases $T_c$ in NbN,[7] VN,[8] and TaN,[9] while in NbC,[10] ZrN,[10] and TaC[9] the value of $T_c$ decreases. Therefore, it is of considerable physical interest to investigate superconductivity at very high pressures in these compounds.

It is well established[11] that the relatively high $T_c$'s in transition metal carbides and nitrides are attributed to a softness of the lattice against hydrostatic deformations, which gives rise to strong electron longitudinal-acoustic coupling. Neutron scattering measurements[12,13] have shown that some superconducting carbides always have anomalies in their phonon dispersion curves, whereas non-superconducting carbides do not exhibit anomalies. Anomalies in the dispersion of the acoustic branches similar to those reported for the superconducting carbides have also been detected in superconducting transition-metal nitrides[14-17] by inelastic neutron scattering as well.

These data also provide important insight into the relationship between superconductivity and lattice instabilities. However, currently the information concerning the change in the characteristic phonon spectra under pressure in these compounds is non-existent, due to the difficulties of neutron scattering experiments, particularly under pressure.

The present investigation was undertaken for purposes of obtaining data on the changes in lattice vibrations under pressure in selected transition-metal nitrides, and to ascertain the contribution of the electron-phonon interaction to different characteristics of the pressure effects on $T_c$ in these materials. We performed the first high-pressure Raman scattering measurements on single crystals of HfN, ZrN, and NbN. High-pressure phonon spectra up to 30 GPa were obtained. The phonon densities obtained from neutron scattering are compared with Raman spectra. We show that the pressure-induced phonon shifts are essential for understanding the pressure effects on $T_c$ in these compounds.

## II. EXPERIMENTAL TECHNIQUES

Single crystals of HfN, ZrN, and NbN were grown by the zone-annealing technique which has been detailed previously.[15-17] Rods of hafnium, zirconium, and niobium with higher than 99.9% purity were zone annealed in a nitrogen atmosphere of nominal 99.99% purity under a pressure of 2 MPa at a temperature regime from 2100 to 3000 °C. Specimens for Raman scattering measurements and structural study were cut from the crystal. Synchrotron X-ray diffraction data collected at Argonne National Laboratory will be described elsewhere.[18]

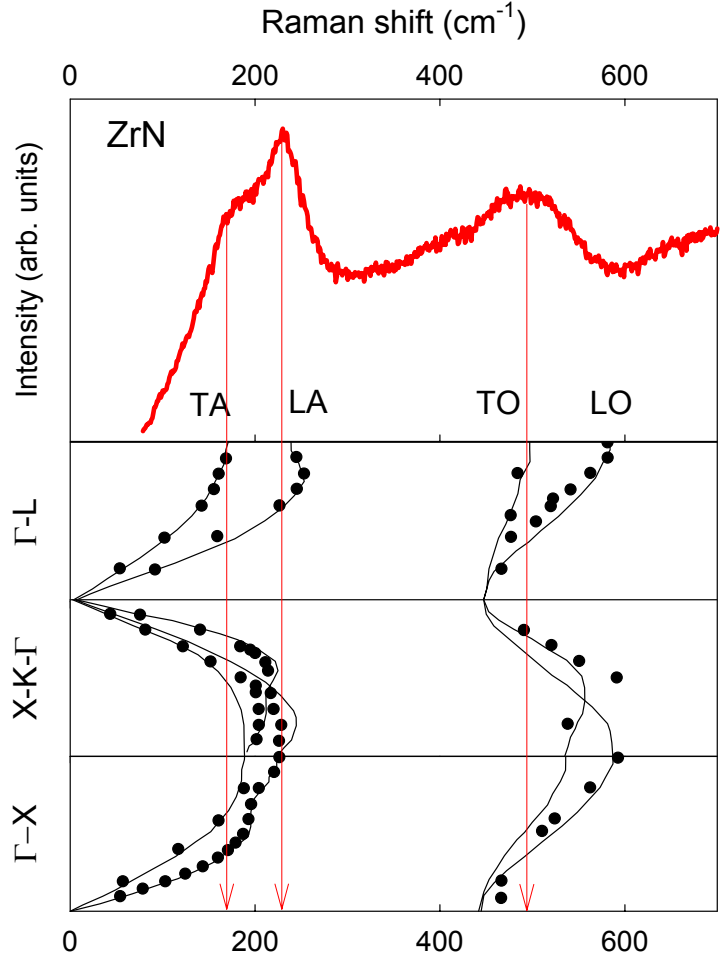

Fig. 1 (color online). Ambient pressure Raman spectra of ZrN (top panel) and neutron inelastic scattering results[15] (lower panels) for the phonon dispersion in ZrN.

Our high pressure Raman system has been described previously.[19,20] For nontransparent and highly reflective samples, a 'quasi-backscattering' (or angular) geometry is employed. This arrangement has the advantage that the specular reflection (or direct laser beam in the case of forward scattering) is directed away from the spectrometer. Compared with the backscattering geometry, this expedient reduces the overall background, and allows the observation of lower-frequency excitations. In the case of metals, use of the angular excitation geometry was found to be crucial.[19] To be compatible with this geometry the diamond seat is modified to allow off-axis entry of the incident light. Specially designed tungsten carbide seats having angular conical holes were used for this purpose. We have used synthetic ultra-pure anvils to reduce diamond fluorescence. Samples were loaded into DAC without any pressure medium and pressure was determined using ruby fluorescence technique. Argon ion laser operating at 514 nm or 488 nm was used for excitation of Raman signals, the laser power on the sample in DAC was about 100 mWt. A typical Raman spectrum from ZrN is compared to the neutron inelastic scattering results[15] in Fig. 1.

### III. PRESSURE EFFECT ON PHONON SPECTRA

Extensive studies[21-23] have proven Raman scattering to be a powerful tool in revealing phonon anomalies in transition-metal nitrides, because of the good agreement between the phonon densities obtained from neutron scattering and the Raman spectra similar to the one-phonon density of states. The stokes sides of the Raman spectrum up to 800 cm$^{-1}$ as a function of pressure for HfN, ZrN, and NbN are shown in Figs. 2-4. The spectrum exhibits two pronounced bands, which are related to the acoustic part and the optical part of the phonon spectrum, respectively. The pressure dependence of the Raman peaks in the low-frequency and high-frequency range is summarized in Table 1.

As can be seen from Fig. 2, transverse and longitudinal branches for Hf vibrations at the BZ boundary (~200 cm$^{-1}$) are well resolved, as well as the higher frequency nitrogen band (~ 500-600 cm$^{-1}$). The low-frequency scattering is caused by acoustic phonons, and the high-frequency scattering is due to optical phonons. With increasing pressure, the two scattering bands shift to higher frequencies. The two peaks in the acoustic range as well as the high-frequency optical branch are in close agreement with the phonon density of states obtained from neutron scattering data in HfN.[16]

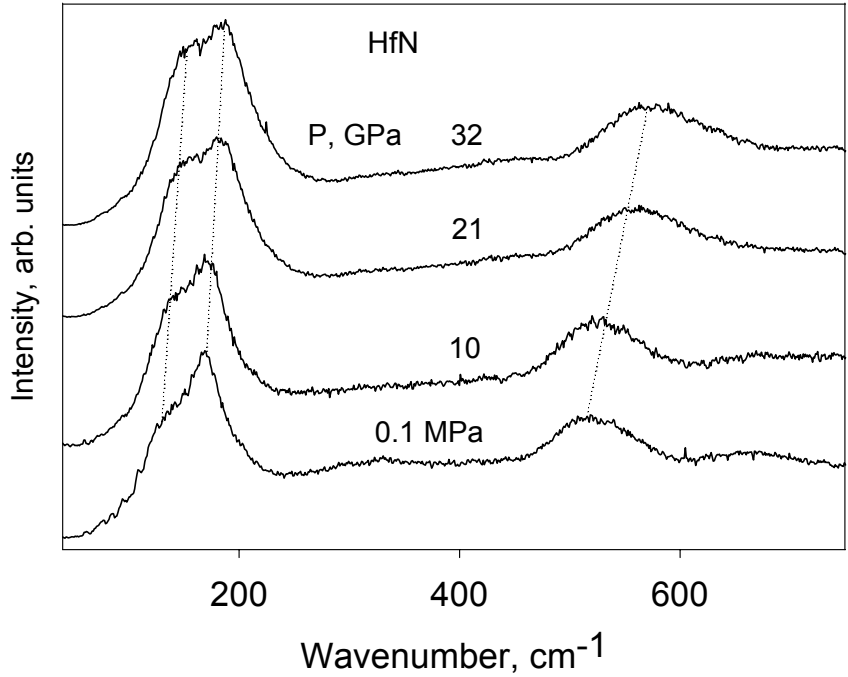

Fig. 2. Raman scattering intensity of a HfN single crystal measured under various pressures.

Similar to HfN, transverse and longitudinal acoustic bands for Zr shown in Fig. 3 are well resolved, however, they are located at higher frequencies since Zr atoms are lighter than Hf atoms. An additional weak band around 700 cm$^{-1}$ may belong to second-order scattering. At ambient condition, the spectrum of ZrN is dominated by strong peaks at 178, 230, and 497 cm$^{-1}$. These characteristics are very similar to the previous Raman measurements.[22] The frequency range of high phonon density of states determined from the coherent inelastic neutron scattering[24] is from 150 to 260 cm$^{-1}$ for the low-frequency acoustic branch and around 500 cm$^{-1}$ for the optical branch, respectively. The application of pressure shifts the peak frequency to higher values. This behavior is similar to the case in HfN shown in Fig. 2. Since the good agreement (see, Fig. 1.) between the Raman spectra and the phonon densities,[24] this shift indicates pressure-induced phonon hardening.

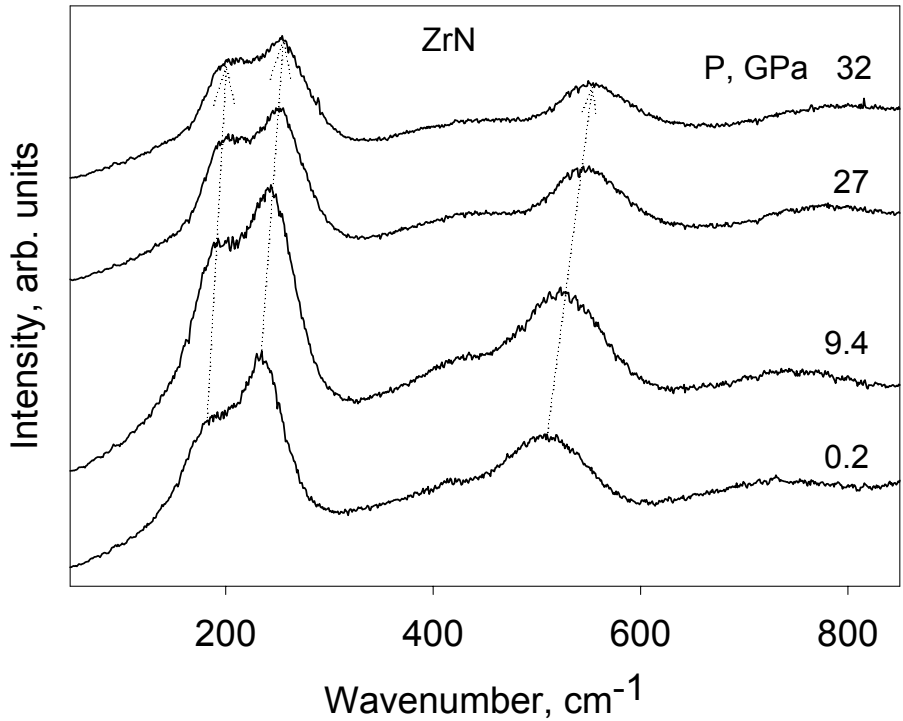

Fig. 3. Raman scattering intensity of a ZrN single crystal measured under various pressures.

Measured high-pressure Raman spectra of NbN are shown in Fig. 4. The high frequency optical branch shows the same overall behavior as HfN and ZrN. Here, transverse and longitudinal Nb bands around 190 cm$^{-1}$ are not well resolved. The relative intensity of the high-frequency (600 cm$^{-1}$) nitrogen band is much lower than in the homologous HfN and ZrN samples. In comparison with the two peaks of the acoustic part observed in HfN and ZrN, the low frequency acoustic part of NbN has two interesting features. First, there is only a broad frequency peak around 190 cm$^{-1}$ in NbN. The relatively broad line shape indicates anomalies in the phonon spectra of NbN. Another prominent feature is the insensitivity of the broad peak to pressure.

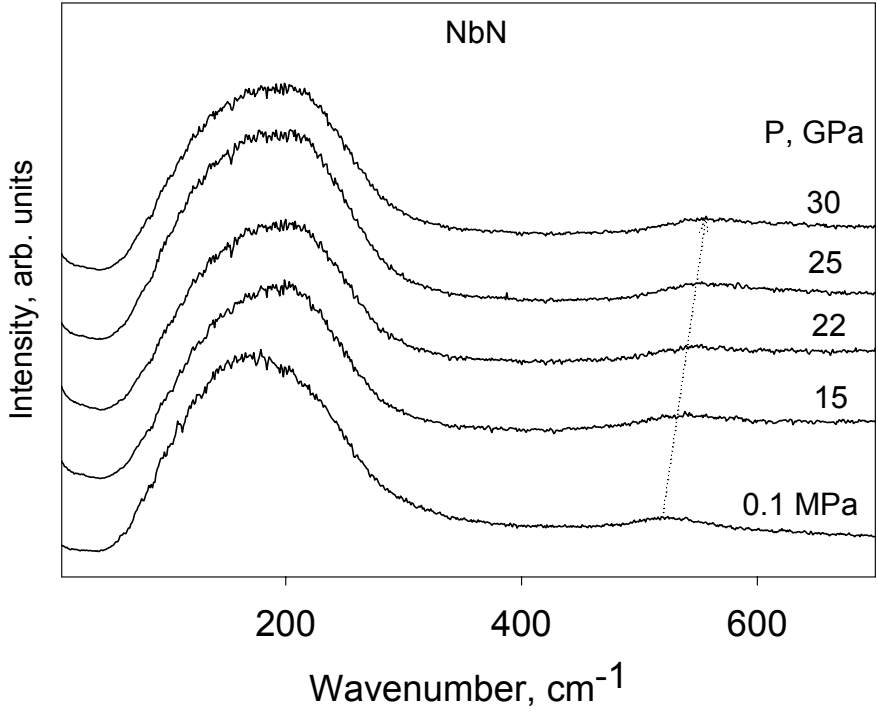

Fig. 4. Raman scattering intensity of a NbN single crystal measured under various pressures.

Information regarding the Raman frequency shift with pressure in these materials should help elucidate the pressure effect on $T_c$. Suppose the two atoms in the unit cell are labeled $T$ and $N$ for transition metal and nitrogen and their masses are represented by $M_T$ and $M_N$, respectively ($M_T \gg M_N$). The product of the mean-square frequencies $<\omega^2>$ and the mass $M$ was obtained from the Raman spectra in terms of the relation[25] $M\langle\omega^2\rangle = M_s\langle\omega_{ac}^2\rangle + M_r\langle\omega_{op}^2\rangle$, where $M_s = M_T + M_N$ and $M_r^{-1} = M_T^{-1} + M_N^{-1}$. Because of the large mass ratio between the transition-metal and the nitrogen atoms, the mass $M_s$ is nearly equal to that of the metal atom and the reduced mass $M_r$ is almost equal to those in nitrogen atoms. In transition-metal nitrides one has[11] $M_T \langle \omega_{ac}^2 \rangle = M_N \langle \omega_{op}^2 \rangle$, as one would expect from nearest-neighbor forces. Therefore, we have $\langle \omega^2 \rangle^{1/2} \approx (2M_T/M)^{1/2} \langle \omega_{ac}^2 \rangle^{1/2} \approx (2M_N/M)^{1/2} \langle \omega_{op}^2 \rangle^{1/2}$ for these materials. This implies that $d \ln \langle \omega^2 \rangle^{1/2} / dP = d \ln \langle \omega_{ac}^2 \rangle^{1/2} / dP = d \ln \langle \omega_{op}^2 \rangle^{1/2} / dP$. It is clear that the pressure-induced phonon frequency shift can be well represented by either the pressure-induced frequency shift of the acoustic branch or the shift of the optical frequency branch.

Table 1. Pressure dependence of the low-frequency (LF) and high-frequency (HF) Raman peaks of the superconducting transition-metal nitrides HfN, ZrN, and NbN.

| HfN | | | | ZrN | | | | NbN | | |
|---|---|---|---|---|---|---|---|---|---|---|
| P (GPa) | LF1 (cm$^{-1}$) | LF2 (cm$^{-1}$) | HF (cm$^{-1}$) | P (GPa) | LF1 (cm$^{-1}$) | LF2 (cm$^{-1}$) | HF (cm$^{-1}$) | P (GPa) | LF (cm$^{-1}$) | HF (cm$^{-1}$) |
| 0 | 125 | 169 | 521 | 0.2 | 179 | 232 | 506 | 0 | 182 | 523 |
| 10 | 136 | 171 | 528 | 9.4 | 189 | 243 | 521 | 15 | 190 | 558 |
| 21 | 146 | 179 | 564 | 27 | 196 | 250 | 547 | 22 | 189 | 571 |
| 32 | 149 | 187 | 576 | 32 | 201 | 255 | 552 | 25 | 189 | 575 |
| | | | | | | | | 30 | 186 | 572 |

A tunneling experiment by Zeller[26] showed that the main contribution to the electron-phonon coupling constant λ comes from the acoustic vibrations that correspond to predominantly metal atom vibrations. Furthermore, Spengler *et al.*[23] reported a close relationship between the first peak of the acoustic branch and the value of $T_c$ in stoichiometric superconducting TiN. Whether or not the frequency of the acoustic part plays an important role in the pressure effect on $T_c$ through the change of λ, it is interesting to see the pressure-induced change of the first peak in the acoustic range. The results from high pressure Raman scattering are shown in Fig. 5. With the application of pressure, the peak frequency initially shifts to higher frequency in all of the nitrides studied. With further pressure increase, the frequency begins to decrease after reaching a maximum at around 15 GPa in NbN. However, pressure always increases the frequency peak in HfN and ZrN in the pressure range up to 32 GPa. Considering that $\lambda \propto \langle \omega_{ac}^2 \rangle^{-1}$ is valid approximately, these observations indicate that $T_c$ would decrease with increasing pressure. Obviously, this behavior is not always true for transition metal nitrides. Thus, we expect that the interplay of the counteracting changes in $\langle \omega_{ac}^2 \rangle$ and the 'electronic' part of $\lambda$ due to pressure is responsible for the pressure dependence of $T_c$.

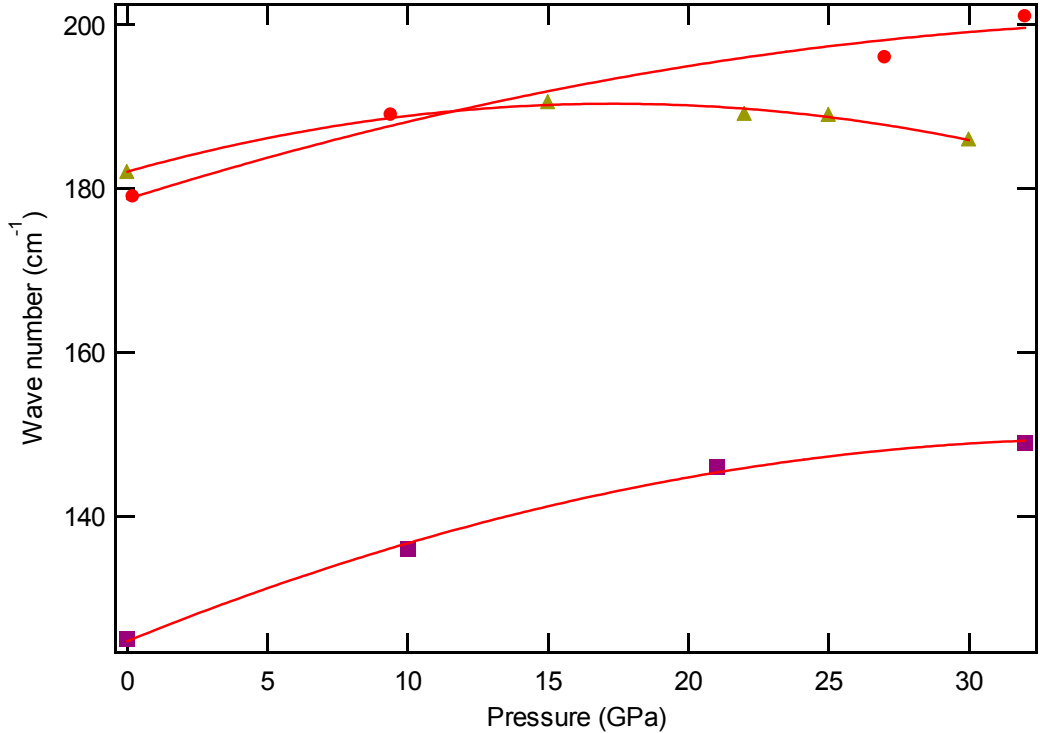

Fig. 5. The pressure dependence of the frequency of the first peak of the acoustic branch in HfZ (squares), ZrN (circles), and NbN (triangles). The lines are guides to the eye.

**IV. PRESSURE EFFECTS ON $T_c$**

For the analysis of the results on the superconducting transition temperature we start with McMillan's $T_c$ formula[27]

$$T_c = \frac{\Theta_D}{1.45} \exp\left[-\frac{1.04(1+\lambda)}{\lambda - \mu^*(1+0.62\lambda)}\right] \quad , \quad (1)$$

which relates $T_c$ to the electron-phonon coupling parameter $\lambda$, the Coulomb repulsion strength $\mu^*$, and a temperature $\Theta_D$ characteristic of the phonons.

Considering the variations of $\Theta_D$, $\lambda$, and $\mu^*$ with pressure or volume and introducing parameters $\varphi = \partial \ln \lambda / \partial \ln V$ and $\xi = \partial \ln \mu^* / \partial \ln V$, we can get the pressure coefficient of $T_c$

$$\frac{d \ln T_c}{dP} = \frac{\gamma_G}{B_0} - \frac{1.04\lambda(1+0.38\mu^*)}{[\lambda - \mu^*(1+0.62\lambda)]^2} \frac{\varphi}{B_0} + \frac{1.04\mu^*(1+\lambda)(1+0.62\lambda)}{[\lambda - \mu^*(1+0.62\lambda)]^2} \frac{\xi}{B_0} \quad , \quad (2)$$

where $B_0 \equiv 1/\kappa_V = -\partial P / \partial \ln V$ is the bulk modulus and $\Theta_D$ is assumed to be proportional to $\langle\omega^2\rangle^{1/2}$ and $\gamma_G = -\partial \ln \langle\omega^2\rangle^{1/2} / \partial \ln V$ being the effective Grüneisen parameter.

The formula for $\mu^*$ due to Morel and Anderson[28] used here is

$$\mu^* = \frac{\mu}{1 + \mu \ln(E_F / \omega_{ph})} \quad , \quad (3)$$

with $\mu = 0.5\ln[(1+a^2)/a^2]$ and $a^2 = \pi e^2 N(E_F)/k_F^2$, where $N(E_F)$ is the electronic density of states at the Fermi energy $E_F$, from which we evaluate the volume dependence of $\mu^*$ as

$$\xi = \mu^* \left[ \frac{2}{3} - \gamma_G - \frac{1-e^{-2\mu}}{2\mu^2}\left(\gamma_N + \frac{2}{3}\right) \right] . \quad (4)$$

Here $\gamma_N = \partial \ln N(E_F)/\partial \ln V$ and the variation of $k_F$ with volume has been calculated from the fundamental definition $k_F = (3\pi^2 Z/V)^{1/3}$ with $Z$ the valency.

The electron-phonon coupling parameter $\lambda$ can be expressed as

$$\lambda = \frac{N(E_F)\langle I^2 \rangle}{M\langle \omega^2 \rangle} \equiv \frac{\eta}{M\langle \omega^2 \rangle} , \quad (5)$$

where $\langle I^2 \rangle$ is the mean-square electron-ion matrix element and $M$ the ionic mass. The McMillan-Hopfield parameter $\eta$ is the electronic part of $\lambda$, which has a local 'chemical' property of an atom in a crystal. Taking the logarithmic volume of Eq. (5), we find

$$\varphi = 2\gamma_G + \frac{\partial \ln \eta}{\partial \ln V} . \quad (6)$$

It has been found[29,30] that the coefficient $\partial \ln \eta / \partial \ln V$ is close to

$$\frac{\partial \ln \eta}{\partial \ln V} = -\gamma_N - \frac{2}{3} . \quad (7)$$

Knowing $B_0$, $\gamma_G$, and $\gamma_N$, one can calculate the pressure derivative of $T_c$ using Eqs. (2), (4), (6), and (7).

We take ZrN and NbN as examples since they have opposite signs of $dT_c/dP$ in order to show how the Raman results provide valuable information on the pressure effect on $T_c$. The structural parameter $B_0$ is taken to be 270 and 350 GPa for ZrN and NbN from the theoretical studies.[31] Such values are consistent with our structural data recently determined at ambient condition and high pressures.[18] As seen in Fig. 5, the lowest frequency of the acoustic phonons initially increases with pressure. If we take the simplifying assumption that the measured lowest frequency peak $\omega_{LF1}$ can be used to replace $\langle\omega^2\rangle^{1/2}$ over the whole frequency range, then we would have the initial value of $d\ln\langle\omega^2\rangle^{1/2}/dP$ of 3.70×10⁻³ and 2.93×10⁻³ GPa⁻¹ for ZrN and NbN, respectively. Thus we have the lattice Grüneisen parameter $\gamma_G$=1.0 for both compounds. Band structure calculations[32,33] show that the electronic density of states $N(E_F)$ initially decreases with pressure at a rate of $d\ln N(E_F)/dP$=-3.60×10⁻³ and -4.49×10⁻³ GPa⁻¹ for ZrN and NbN, respectively. Combining this calculated value and $B_0$, we get $\gamma_N$=0.97 for ZrN and 1.57 for NbN. Using Eqs. (6) and (7), we then obtain the value of $\varphi$ of 0.36 and -0.24 for ZrN and NbN, respectively. The volume dependence of $\mu^*$ can be derived from Eq. (4) once having the values of $\gamma_G$, $\gamma_N$, $\mu^*$, and $\mu$. Since $a^2$ has a typical value of 0.4,[28] we have $\mu$=0.63. Studies of the isotopic mass dependence of $T_c$ and the inversion of superconducting tunneling curves have revealed that $\mu^*$ values all lay in the range between 0.1 and 0.2. The Coulomb pseudopotential $\mu^*$ was determined to be 0.11 for ZrN and 0.13 for NbN.[34] Based on these parameters, we obtain $\xi$=-0.20 for ZrN and -0.30 for NbN.

The heat capacity measurements yield the Debye temperature $\Theta_D$ of 515 K for ZrN and 363 K for NbN.[35] Substituting the experimentally determined $T_c$'s of 10.00 and 14.94 K gives the electron-phonon coupling constant $\lambda$ of 0.6266 and 0.906 for ZrN and NbN, respectively. With the parameters determined above, we have calculated the pressure derivative of $T_c$ by using Eq. (2). The calculated $dT_c/dP$ is -0.012 and 0.05 K/GPa for ZrN and NbN, respectively. The sign of our calculated $dT_c/dP$ clearly agrees with the experimental data.[7,10] Therefore, Raman scattering under high pressure indeed provides essential information for understanding of the pressure effect on $T_c$. The calculated $dT_c/dP$ of 0.05 K/GPa for NbN is in excellent agreement with the measured value of 0.04 K/GPa.[7] For ZrN, Smith[10] reported a $dT_c/dP$ of -0.17 K/GPa from high pressure measurements. Our theoretical result is one order smaller in magnitude than the experimental result. This is not surprising since $T_c$ depends significantly on the nitrogen composition. It has been found[9] that the higher the $T_c$ value, the smaller its pressure derivative in magnitude. The small absolute value of $dT_c/dP$ by using the estimated $\gamma_G$ obtained from our measurements implies that our sample has a higher $T_c$ than Smith's sample.

We would like to emphasize that the opposite sign of $dT_c/dP$ observed for ZrN and NbN results from the competition between the electronic and phonon part of λ. We found that the contribution to $dT_c/dP$ is mainly determined by the sum of the first two terms on the right of Eq. (2). If the contribution from the electronic part of λ outweighs the change in $\langle\omega^2\rangle$, we would have a negative $\varphi$ and a positive $dT_c/dP$, accordingly. Therefore, the pressure derivative of $T_c$ for NbN mainly originates from pressure-induced changes in the electronic part of the electron-phonon coupling parameter. In contrast, the pressure-induced phonon frequency shifts dominate the pressure effect on $T_c$ in ZrN.

The expressions for $\gamma_G$, $\varphi$, and $\xi$ can be integrated to give $\Theta_D(V)=\Theta_D(0)[V/V_0]^{-\gamma_G}$, $\lambda(V)=\lambda(0)[V/V_0]^{\varphi}$, and $\mu^*(V)=\mu^*(0)[V/V_0]^{\xi}$, respectively. Here $V$ and $V_0$ are the unit cell volumes under the applied pressure and at ambient pressure, respectively. These two volumes can be related according to the first-order Murnaghan equation of state $V(P)=V(0)\left(1+B_0'P/B_0\right)^{-1/B_0'}$, where $B_0'$ is the pressure derivative of the bulk modulus. Thus, one has the pressure dependence of $\Theta_D$, λ, and $\mu^*$

$$\Theta_D(P) = \Theta_D(0)\left[1 + \frac{B_0' P}{B_0}\right]^{\gamma_G / B_0'}$$

$$\lambda(P) = \lambda(0)\left[1 + \frac{B_0' P}{B_0}\right]^{-\varphi / B_0'} \qquad (8)$$

$$\mu^*(P) = \mu^*(0)\left[1 + \frac{B_0' P}{B_0}\right]^{-\xi / B_0'}$$

The pressure dependence of $T_c$ is then readily obtained if one substitutes Eq. (8) into Eq. (1).

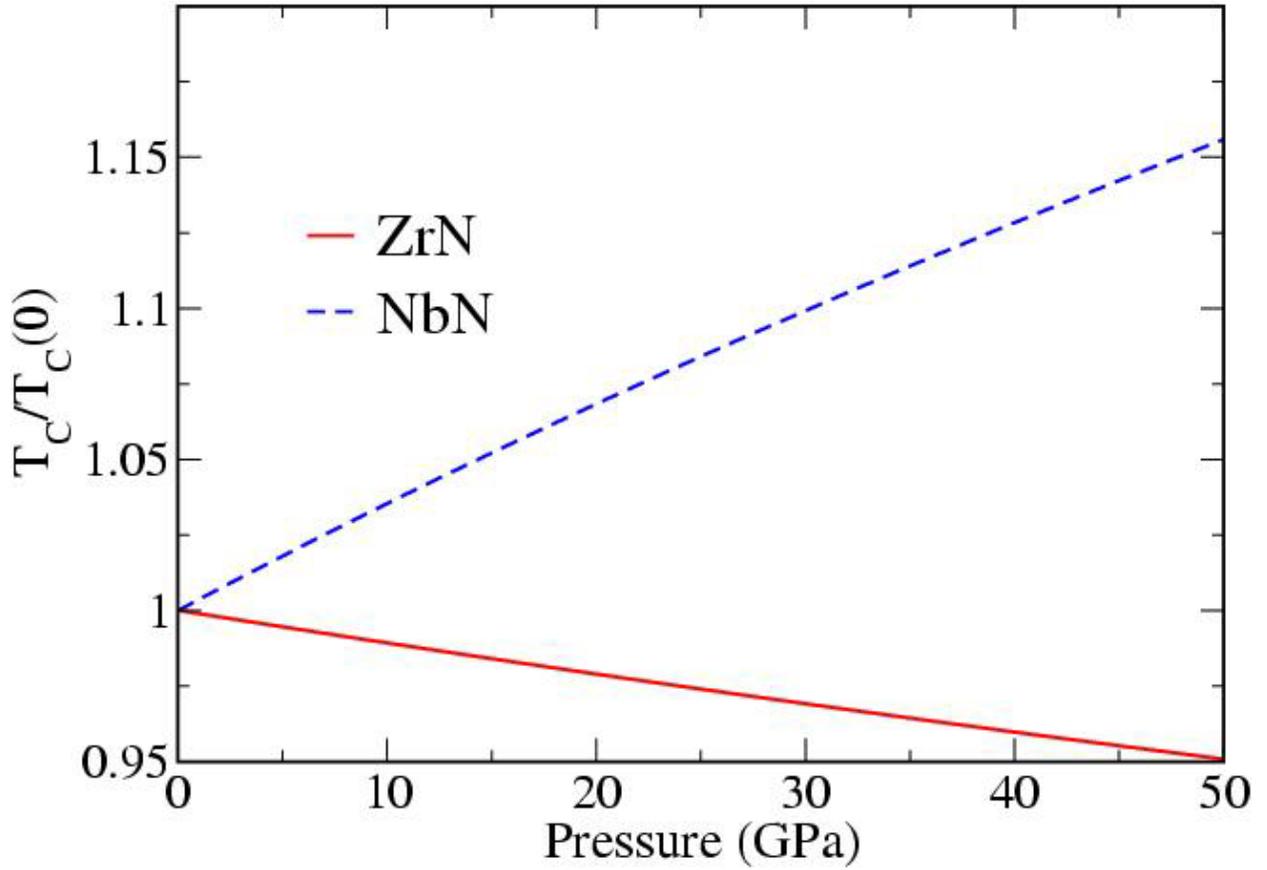

Fig. 6. Pressure dependence of the normalized superconducting transition temperature in ZrN and NbN up to 50 GPa.

As shown by Eq. (8), we need the value of $B_0'$ in order to investigate the high pressure behavior of $T_c$. Generally, it is reasonable to take $B_0'=4$ for the NaCl-type structure.[36] The theoretical results for the variation of the normalized $T_c$ as a function of pressure up to 50 GPa for ZrN and NbN are shown in Fig. 6. We found that $T_c$ decreases with pressure and remains 9.5 K at even 50 GPa in ZrN, while an increase of $T_c$ on compression is observed for NbN. Direct high pressure measurements of $T_c$ are required to examine our prediction.

## V. CONCLUSIONS

High-pressure Raman scattering spectra of transition-metal nitrides HfN, ZrN, and NbN exhibit two pronounced bands, which are related to the acoustic part around 200 cm$^{-1}$ and the optical part around 550 cm$^{-1}$ of the phonon spectrum. Transverse and longitudinal acoustic bands for Hf and Zr are well resolved, while the two bands in Nb are broad and overlapped. These characteristics are supported by the phonon density of states determined from the inelastic neutron scattering. Application of pressure shifts the peak frequency to higher values. Due to the good agreement between the Raman spectra and the phonon densities, this shift indicates pressure-induced phonon hardening.

Based on the phonon frequency shifts with pressure as well as the related parameters determined theoretically, we calculated the pressure derivative of $T_c$ and the variation of $T_c$ with pressure for ZrN and NbN using the McMillan formula. The calculated $dT_c/dP$ for NbN is in excellent agreement with experiments. Our calculated $dT_c/dP$ for ZrN is one order smaller in magnitude than the experimental result, indicating a relatively high $T_c$ of our sample. The variation of $T_c$ with pressure up to 50 GPa is predicted for ZrN and NbN. We have demonstrated that pressure-induced phonon frequency shifts and changes in the electronic part of the electron-phonon coupling parameter are mainly responsible for the pressure effects on $T_c$ in ZrN and NbN, respectively. Extended measurements of the pressure dependence of $T_c$ in transition-metal nitrides are in progress.

## ACKNOWLEDGMENTS

This work was supported by the U.S. Department of Energy under awards DEFG02-02ER4595 and DEFC03-03NA00144.